# GENERATIVE AI IN SELF-DIRECTED LEARNING: A SCOPING REVIEW

Jasper Roe [1*], Mike Perkins [2]

[1] James Cook University Singapore, Singapore.
[2] British University Vietnam, Vietnam.

* Corresponding Author: jasper.roe@jcu.edu.au

## Abstract

This scoping review examines the current body of knowledge at the intersection of Generative Artificial Intelligence (GenAI) and Self-Directed Learning (SDL). By synthesising the findings from 18 studies published from 2020 to 2024 and following the PRISMA-SCR guidelines for scoping reviews, we developed four key themes. This includes GenAI as a Potential Enhancement for SDL, The Educator as a GenAI Guide, Personalisation of Learning, and Approaching with Caution. Our findings suggest that GenAI tools, including ChatGPT and other Large Language Models (LLMs) show promise in potentially supporting SDL through on-demand, personalised assistance.

At the same time, the literature emphasises that educators are as important and central to the learning process as ever before, although their role may continue to shift as technologies develop. Our review reveals that there are still significant gaps in understanding the long-term impacts of GenAI on SDL outcomes, and there is a further need for longitudinal empirical studies that explore not only text-based chatbots but also emerging multimodal applications.

# Introduction

Since November 2022, the development of Generative AI (GenAI) has proceeded at a startling speed (Rudolph et al., 2023b), outpacing the implementation of social and legal frameworks (Khosravi et al., 2022). Whether this development will continue at such a scale and velocity in the future is unknown, but the full impact of technology available today is yet to be fully understood. In the field of education, GenAI remains an immature field of research (Chiu, 2023), yet it is likely to define educational scholarship in the years to come. The promise of personalised learning and intelligent tutoring systems are among the most provocative and exciting opportunities offered by these new technologies, yet at the same time, concerns abound regarding tools such as ChatGPT producing 'nonsense' (Rudolph et al., 2023a) or 'botshit' (Hannigan et al., 2024), and GenAI tools such as ChatGPT have been critiqued as 'bullshitters' (Hicks et al., 2024) rather than useful tools which offer genuine value to society. From the perspective of critics, the benefits of GenAI are unjustifiable, given the extreme resource use required to train new models (Bourzac, 2024) and potentially huge planetary costs of AI development (Driessens & Pischetola, 2024).

On the other side of this argument, in education at least, empirical research has shown that GenAI tools can positively impact academic achievement (Kestin et al., 2024; Sun and Zhou, 2024; Wu & Yu, 2024). Despite GenAI's tendencies to produce incorrect and counterfactual output confidently, GenAI tools may be able to help lessen foreign language learner anxiety (Zheng, 2024), enhance learning experiences by diversifying learning materials, and offering scalable personalised support (Yan et al., 2024), as well as improve the teaching of academic writing (Roe, O'Sullivan, et al., 2024), simplify research processes (O'Dea, 2024), and complement existing research techniques and methodologies (Perkins & Roe, 2024). Institutions are now exploring ways to implement the acceptable use of GenAI in ethical ways in educational assessment (Furze et al., 2024; Perkins, Furze, et al., 2024), recognising the potential benefits, inadequacy, and unenforceability of outright bans or prohibitions of publicly available technology. However, concerns remain regarding ethics, academic integrity, equity, human values and human rights (Cotton et al., 2023; Eaton, 2024; Perkins, 2023; Rudolph et al., 2024). To address such concerns, increased focus is now being given to ways to develop learner skills and re-evaluate assessments; for example, by cultivating self-regulated learning skills for students (Xia et al., 2024), developing evaluative judgement among students (Bearman et al., 2024), and focusing on assessment validity rather than cheating (Dawson et al., 2024).

Against this increasingly polarised backdrop, there remains a critical need to systematically examine the potential impacts of GenAI on different modes of education. Self-Directed Learning (SDL) is an area in which these technologies' capabilities for personalised, on-demand support appear particularly relevant. With this in mind, we contribute to the literature by conducting an initial scoping review of the current research on Generative AI and Self-Directed Learning. This review has three purposes:

1. To map the emerging landscape of GenAI applications in SDL contexts.
2. To critically evaluate whether current evidence supports GenAI as a net benefit for SDL when considering both opportunities and risks.
3. To identify key research gaps and future directions that could inform a more effective integration of GenAI into SDL practices.

By synthesising the current state of knowledge, this review provides a foundation for understanding how GenAI might impact SDL and help set the direction for future research in this area.

## Self-Directed Learning and GenAI

SDL refers to learning in which the learner directs the conceptualisation, design, conduct, and evaluation of a learning project. However, SDL does not have to be totally individual and can involve collaborative and group learning (Brookfield, 2009), or being part of a project or other group activities (van Woezik et al., 2021). Scholarship on SDL began with the seminal work of Tough (1967), and was built upon by pioneers such as Knowles (1975), who defined eight elements that constitute SDL, and Candy (1991), who identified four major domains that comprise SDL. From a theoretical perspective, SDL can be viewed from a collaborative-constructivist theory of education (Garrison, 1997), and a full SDL programme requires learners to identify their learning needs, determine their objectives, decide on evaluation, identify suitable resources, and evaluate the end product of their learning (Iwasiw, 1987).

The positive impact of SDL includes the fact that it can help learners become more responsible for the decisions associated with their education (Hiemstra, 1994). SDL competence is positively related to constructs such as performance, aspiration, creativity, curiosity, and life satisfaction (Boyer et al., 2014). SDL has been said to have 'unlimited potential' to enhance the success of adult learners (Brockett & Hiemstra, 1991) and is particularly important for this cohort as it enables them to balance learning with other commitments (Khiat, 2017). This said, there is no common consensus on approaches to implementing SDL (O'Shea, 2003), and many related concepts, such as self-planned or autonomous learning, are used synonymously with SDL, although they have different emphases (Hiemstra, 1994). Consequently, it has been argued that SDL is often mistakenly assigned a shallower meaning, such as self-study (Silén & Uhlin, 2008). SDL has also been critiqued for being culture-bound and essentially aligned with the Western cultural tradition of individualism (Brookfield, 2009), although even in a Western context, research has shown a lack of instructor support for implementing SDL (Wilcox, 1996). To some extent, this is in line with current criticisms of the culture-bound nature of GenAI models, which reflects a predominantly Western-centric worldview as a result of their training data (Roe, 2024).

A strong relationship exists between technology and SDL. Digital technologies can support the implementation or facilitation of SDL (Godsk & Møller, 2024; Morris & Rohs, 2023) and well-designed online learning environments can benefit SDL opportunities for learners (Sumuer, 2018). Additionally, technology use has a positive effect on SDL and student engagement (Rashid & Asghar, 2016), and students who have a greater acceptance of technology have shown a higher attitude towards technology-based SDL (Pan, 2020). However, technology-mediated SDL requires additional teacher attention (Boyer et al., 2014; Monib et al., 2024) as students may not be competent in using new technologies (Morris & Rohs, 2023). Such issues can easily be translated into the use of GenAI, the limitations of which include access and user competence, as well as concerns over accessibility (Nyaaba et al., 2024; Perkins, Roe, et al., 2024). Given the fact that GenAI models, such as ChatGPT, Claude, or Gemini, can provide instantaneous, convincing (though not always accurate) information, and that GenAI tools may act to an extent similar to intelligent personal tutors or virtual peers (Kim et al., 2024), it is possible that these technologies could help develop SDL competencies, making the necessity for a scoping review all the more compelling.

## Scoping Review Procedure

As SDL and GenAI is an emerging area of study, we opted to use a scoping review procedure in line with the Preferred Reporting Items for Systematic Reviews and Meta-Analyses (PRISMA) scoping review extension (PRISMA-SCR) developed by Tricco et al. (2018). According to Tricco et al. (2016), such reviews are useful for outlining the concepts of a specified area of research and for understanding the available research base. At the same time, scoping reviews can be useful for identifying knowledge gaps and examining emerging unclear evidence (Munn et al., 2018).

We searched multiple academic databases including SCOPUS, Google Scholar, ERIC, and Web of Science. Our rationale for these databases was to cover both established sources of academic knowledge and, given the emerging nature of the topic, cast a wide net which included non-traditional source material. We decided to include Google Scholar, as research suggests that it can be a powerful complement to existing traditional search methods and contains relevant grey literature (Haddaway et al., 2015). We set the inclusion criteria as publications beginning from 2020 to 2024, in line with the public release of one of the most noteworthy GenAI models of the present moment: GPT-3 by OpenAI. We also focused specifically on Self-Directed Learning and GenAI, with any methodology or topic included, and in English. We decided to include grey literature, conference proceedings, journal articles, and editorials. The search was conducted by two reviewers using the criteria shown in Table 1.

**Table 1: Inclusion Criteria**

| Inclusion Criteria | Description |
| --- | --- |
| Time of Publication | Published between 2020 – 2024 |
| Language | English |
| Focus | Self-Directed Learning and Generative AI |
| Databases used | SCOPUS, Google Scholar, ERIC, and Web of Science. |
| Type of Publication | Peer reviewed articles, conference papers, preprints, and relevant grey literature (e.g. reports from reputable organisations). |
| Search Terms | Self-Directed Learning, SDL, Generative Artificial Intelligence, GenAI |

## Results

We screened the titles and abstracts of 24 studies. Following this, we read the papers in their entirety. Given the limited number of publications and the nature of our scoping review, we decided not to use formal or quantitative methods of analysing reviewer agreement and instead relied on consensus building through discussion and collaborative reading. After a full reading of the papers, we removed three duplicates and two irrelevant studies, resulting in 18 studies which met our inclusion criteria. This included 15 empirical studies in peer-reviewed journals, one empirical study in conference proceedings, one theoretical journal article, and one theoretical book chapter. One output was produced in 2022, three in 2023, and 14 in 2024. We then charted the data and noted patterns in temporal distribution, publication type, and content. A summary of the identified studies is presented in Table 2.

**Table 2 – Data Charting**

| # | Author | Type | Year | Type | Outcome |
|---|---|---|---|---|---|
| 1 | Lashari & Umrani | Journal | 2023 | Empirical research | There is a need for additional research in academia regarding how ChatGPT can benefit outcomes in second language learning. |
| 2 | Yildrim et al. | Book Chapter | 2024 | Theoretical | Despite AI making life easier for educators, it is important that the educator contributes to personalizing student learning. |
| 3 | Li et al. | Conference Proceedings | 2024 | Empirical Research | ChatGPT shows potential as a tool for assisting with self-directed language learning technologies. |
| 4 | Wang et al. | Journal | 2024 | Empirical Research | AI technologies should be used to empower students as active participants in SDL to promote understanding, personalization, reflection and collaboration. AI may be able to lead to new learning strategies and benefit motivation. |
| 5 | Wu et al. | Journal | 2024 | Empirical research | Educators should design activities using AI in alignment with clear guidelines and offer opportunities to critically evaluate GenAI content. This, with additional external feedback and resources, can enhance SDL learning with AI. |
| 6 | Jayasinghe | Journal | 2024 | Empirical research | Studies of senior management in software industries suggest that GenAI is viewed as positive for SDL opportunities. |
| 7 | Lin | Journal | 2024 | Empirical research | The use of AI and ChatGPT can assist adult learners to engage in SDL by enabling them to set learning goals, locate resources, construct plans for learning, and provide monitoring on performance. |
| 8 | Shalong et al. | Journal | 2023 | Empirical research | A custom trained GPT (LearnGuide) is able to enhance SDL and critical thinking experiences among medical students. |
| 9 | Ouaazki et al. | Conference Proceedings | 2024 | Empirical research | Using GenAI in conversational interactions can support SDL, especially when prompt-tuning is used, although a consequence may be hindering ease of use. Risks include diminished utility over time and overreliance. |
| 10 | Chang et al. | Journal | 2024 | Empirical research | ChatGPT can be used to support SDL among nursing students, and customisation could help provide optimal conditions. |
| 11 | Bosch & Kruger | Journal | 2024 | Theoretical article | The use of AI chatbots can promote SDL and student autonomy through various techniques, yet the potential of such tools in empowering learners is underexplored. |
| 12 | Indriani et al. | Journal | 2024 | Empirical research | The use of ChatGPT significantly and positively affected SDL among learners across multiple disciplines in undergraduate and postgraduate cohorts. |
| 13 | Han et al. | Journal | 2022 | Empirical research | GenAI chatbots can be used as educational assistants to promote nursing students' interest in SDL, and also improve non-face-to-face communication skills. |
| 14 | Li et al. | Journal | 2024 | Empirical research | ChatGPT can enhance learner growth by generating content which has contextual meaning for learners. Educators must take a critical role when considering how to integrate GenAI in SDL. |
| 15 | Sadiq et al. | Journal | 2024 | Empirical research | Using ChatGPT among education students improved autonomous learning, motivation, engagement, and SDL skills. had a significant impact on student research skills, although students with high levels of autonomous motivation are likely to benefit the most from using ChatGPT in research skill development. |
| 16 | Dizon | Journal | 2024 | Empirical research | Participants in this research viewed using ChatGPT for positively but showed scepticism on the use of AI for language learning. |
| 17 | Askarbekuly & Aničić | Journal | 2024 | Empirical research | The use of ChatGPT to develop assessment questions for SDL purposes was viewed generally positively, although issues included the data quality, alignment of LLM and instructor, and the user interface. |
| 18 | Ali et al. | Journal | 2023 | Empirical research | The use of a custom GenAI chatbot assisted teachers effectively in facilitating SDL among K-12 students. |

To identify the themes presented in the literature, we followed Braun and Clarke's (2006, 2019, 2023) six-step approach to reflexive thematic analysis, focusing on our centrality and position as researchers in the data interpretation and theme development process and avoiding a search for an objective set of emergent themes. We chose to use an inductive approach, given that such usage is suitable for exploring new terrain (Clarke & Braun, 2017). We began by closely reading the selected papers before conducting open coding and identifying patterns of recurrent or shared meanings. Following the six-step process, we collaboratively generated a list of initial themes and reflected on our own experiences as educators and researchers in understanding the data. Through an iterative process, we reviewed and defined themes before naming them and producing our report. Figure 3 illustrates the themes developed from the literature review.

**Table 3 – Theme Identification**

| Theme | Description |
| --- | --- |
| 1. GenAI as a Potential Enhancement for SDL | This theme refers to both the theoretical and realised benefits of GenAI in SDL in aiding student learning, teaching, or enhancing the learning process. |
| 2. The Educator as a GenAI Guide | This theme refers to the concept that the role of an instructor or teacher must change significantly yet retains a pivotal degree of importance in the SDL process as a guide to GenAI use. |
| 3. Personalisation of Learning | This theme refers to the potential for GenAI to provide content, support, and information that is tailored to the learner's preferences or needs, thus considered 'personalized'. |
| 4. Approaching with Caution | This theme relates to the highlighting of potential negatives relating to the use of GenAI in SDL, including risks to the learner. |

**Theme 1: GenAI as a Potential Enhancement for SDL**

A recurrent topic that we noted throughout the literature is the possibility of GenAI tools, such as ChatGPT, to support and enhance learning in an SDL context. Often, this topic was raised not as the result of an empirical study but was theoretically postulated as one of the potential benefits. Lin (2024) for example, describes the way that traditional SDL activities, such as seeking resources, setting learning goals, and designing learning plans may be assisted by ChatGPT, while Li et al. (2024) identify that SDL is a context in which learners can explore tools such as ChatGPT, taking an experimental approach and identifying whether it may help in achieving their educational goals. Similarly, Wu et al. (2024) noted that using cutting-edge models may further extend these enhancements, enabling the creation of multimodal resources for learning, real-time continuous feedback, and evaluation. From our own experience as researchers, we note that there is data suggesting that feedback must include a human element to receive acceptance from students and teachers (Roe, Perkins, & Ruelle, 2024). Wang et al. (2024) similarly highlight that cognitive and affective benefits in using GenAI for SDL are appealing, drawing on empirical data from students which suggests that by using ChatGPT for brainstorming and ideation, it enabled learners to spend more time engaged in the writing and revising process.

Simultaneously, some empirical studies have demonstrated the enhancing effects of GenAI. Data from Indriani et al. (2024) supported the beneficial effects of using ChatGPT in an SDL program and found a positive relationship between ChatGPT use and learning motivation in SDL. Han et al. (2022) found that the use of an AI chatbot program could promote interest in SDL among learning students and that using an adaptive learning system improved SDL capabilities among mathematics learners. Shalong et al. (2024) found through a 14-week randomised controlled trial in which students using a custom AI facilitation tool improved SDL among medical students. In summary, a significant focus in the literature projects a positive and forward-looking perspective, in which GenAI enhances SDL. In engaging reflexive thinking, as education scholars are preoccupied with new technologies, we tried to view this theme from a critical perspective and contend that while the initial data is compelling, much of the perceived enhancement remains at the theoretical stage.

**Theme 2: The Educator as a GenAI Guide**

Whether the development of advanced GenAI tools will lead to a decreased need for teachers, or whether the role of an instructor will change, has been one of the key debates regarding the current AI moment and education, and in thinking reflexively, we approached this topic with some critical scepticism. In developing this theme, we draw on a recurring pattern of the changing role of the teacher in the literature, noting that, in relation to SDL and GenAI, there is a focus on the centrality of the teacher in supporting the adaptation of these technologies to SDL practices, although at the same time, the role of the teacher may be expected to shift and change in multiple ways. Yildrim et al. (2023) for example, claim that in a GenAI enhanced SDL protocol, the teacher will need to relinquish authority and take a facilitative role, working with AI tools, while also receiving training on how to guide students to establish their own learning goals and locate resources while using AI tools. The authors also assert that in a GenAI-enabled SDL environment, the teacher will need to understand each student's capabilities, and that AI tools are not yet able to capture this, thus suggesting a shifted, yet not redundant, role of the teacher. Similarly, Wang et al. (2024) envision a future in which educators play a leading role in guiding students to experiment with AI tools, while also tailoring activities to develop critical AI literacy in learners and help them overcome any potential challenges that stand as an impediment to following an SDL protocol using GenAI.

This is also mentioned by Bosch and Kruger (2024), who highlight that educators may face challenges in integrating GenAI chatbots into SDL, which can occur as a result of misalignment with pedagogical aims or a lack of technological proficiency for both students and teachers. Similarly, under this theme, the educator's role as a guide to help students learn how to use GenAI is foregrounded. For example, Li et al. (2024), highlight that in their study, teachers showed students how to use ChatGPT to address course content, and thus suggest that educators train students to solve problems they encounter independently, while Li, Wang et al. (2024) highlight that in an SDL context, the teacher's role may be to cultivate self-discipline in conducting SDL with AI tools.

**Theme 3: Personalisation of Learning**

The third theme intersects with the shifting role of the educator and the potential benefits of GenAI in SDL but refers to a larger transformation over the process of learning. Throughout our analysis, we noted that the literature tended to point towards an expected expansion of GenAI capabilities. For example, Lashari and Umrani (2023) argue that the role of ChatGPT is likely to grow in the field of learning, while Ali et al. (2023) contend that the use of a GPT-4 based chatbot offers a more personalised, interactive learning environment that mimics everyday interactions, hinting at a significant shift in how learning takes place. Outside of higher education, initial research on how business leaders view the potential of GenAI-enabled SDL suggests that it could be a method of driving innovation and advantage in the commercial stage (Jayasinghe, 2024).

The concept of the personalisation of learning has recurred in several instances. Wu et al. (2024) specifically, describe the traditional lack of real-time interaction in SDL from teachers and peers as somewhat mediated by the interactive nature of GenAI tools, while Ali et al.

(2023) point out the use of customised training, emojis, and in future, multimedia, helps to create a close communicative environment. Dizon (2024) notes that for SDL language learning, using GenAI, personalised learning emerged as a key theme among respondents, suggesting that it held value in using ChatGPT for language learning tasks, while Li et al. (2024) highlight that the functionality of ChatGPT and its advances in the future may lead to the generation of more personalised SDL tasks. Therefore, although AI makes some parts of the job easier for the educator, the educator is still an important part of the process of personalising learning for each student. (Yildirim et al., 2023).

**Theme 4: Approaching with Caution**

Finally, although much of the literature that we surveyed suggested a highly optimistic view of the potential for SDL to be made more effective through the integration of GenAI, consistent concerns were found on some topics. For example, Lin (2024) highlights that one of the challenges in integrating ChatGPT into SDL is ensuring that learners do not become over reliant on using the technology. Therefore, it is important for educational institutions to provide frameworks that support a limited use. Furthermore, Lin (2024) points out that ChatGPT is limited by its accuracy and tendency to provide false or irrelevant information. A similar concern is given by Lashari and Umrani (2023), who contend that with ChatGPT specifically, responses are not generated from an up-to-date store of knowledge, which presents a problem in SDL usage, especially in disciplines which are quickly developing new knowledge. Similarly, Wang et al. (2024) point out that an overreliance on AI is a potential challenge for students engaged in using AI tools for SDL. Other challenges that came under this theme include the unknown impact of manipulating AI tools, given that Ouaazki et al. (2024) found that prompt-tuning ChatGPT for student usage led to decreased perceptions of ease of use and a potential decline in students' attitude towards the technology. Finally, challenges noting the lack of factual integrity, and concerns over current GenAI tools generating monotonous, repetitive outputs is noted by Li et al. (2024), which impacts the possibility of providing rich and deep learning experiences in language learning through SDL.

## Discussion

The findings of this review provide significant insights into the emerging research base of SDL and GenAI. There is no doubt a growing focus on how these new technologies can benefit SDL, which is well-aligned with SDL's historical affinity with technology. In reviewing the literature and engaging in an inductive thematic analysis, we noted that almost all the sources we investigated promoted the potential benefits of GenAI in benefiting learners in SDL as a result of multiple factors, including flexibility, agility, personalisation, and the abilities of GenAI tools themselves. The term 'GenAI' was solely used to refer to Large Language Models (LLMs) such as ChatGPT, with no published studies considering the broader possibilities of GenAI tools to generate imagery, video, or audio. This suggests that to date, there has been a major focus on the use of chatbot GenAI tools to support SDL, and the literature tends towards a focus on personalised tutoring. As multimodal GenAI continues to develop, the benefits that it may bring to SDL could also be explored, such as the use of NotebookLM (Google, 2024) to generate podcasts and topic summaries to benefit SDL, the creation of images and diagrams to support comprehension of topics, or the use of synthetic media and deepfake technology, which is poised to become integrated into educational practices (Roe, Perkins, & Furze, 2024).

We found that much of the literature speaks to the importance of educators in GenAI-enabled SDL. Although it may be expected that the emergence of GenAI tools would lead to a diminished role for teachers, the literature seems to have recurrent shared framing of the teacher as the cornerstone of the learning process. We noted a strong emphasis on the ideal that GenAI tools should not be viewed as a replacement for a teacher and that SDL still requires an instructor's presence to assist with many aspects of the learning process. At the same time, some sources have suggested that the role of the teacher may retain importance but simultaneously transform in an SDL environment, taking more responsibility for developing critical AI literacy in students and enabling them to make use of the affordances offered by GenAI in completing SDL activities. The findings in our review challenge the popular narrative of AI replacing educators, at least in SDL contexts, and offer insights as to where the development of a teacher-AI relationship is heading in supporting students to achieve learning outcomes.

Personalisation as a benefit recurred throughout our analysis of the literature, taking on many different forms, including personalisation of learning outcomes, learning material, assessment, and feedback. SDL in a traditional asynchronous environment may constrain opportunities for personalised interaction and support from teachers or peers. In online learning, especially when learning is self-directed, there are fewer opportunities to receive instant help (Broadbent & Lodge 2021). In the literature, there appears to be an underlying assumption that this functionality may be replaceable with 24/7 learning support from a GenAI tool. At the same time, this has to be counterbalanced with a common risk that was mentioned throughout the material: the potential for students to become overreliant on such tools in an SDL environment. A critical analysis, or further research on how effectively current GenAI tools can replicate these authentic social interactions and whether there is potential for harm in doing so, is therefore an urgent research task.

In drawing these threads together, the current research base on GenAI in SDL suggests that while the future is promising for the potential of these tools to benefit learners and teachers (albeit with great transformations to the SDL process), there are multiple pitfalls to be aware of. These mainly relate to overreliance on GenAI tools, a lack of critical AI literacy, and the fact that GenAI can sometimes produce incoherent, incorrect, or counterfactual information, while also not remaining perpetually up-to-date as a result of the way in which the models are trained on preexisting data. Our review reveals several priority areas for future research and development. Longitudinal studies are needed to assess the long-term impact of the GenAI on SDL outcomes across multiple educational and cultural contexts. Further, more research is needed to understand how a broader range of GenAI tools may be implemented in SDL practices, rather than just LLM chatbots such as ChatGPT. The assumption that GenAI can act as a personal tutor, virtual peer, or helper warrants immediate investigation if it is to be deployed at a scale in SDL environments, and more studies are needed to understand how to foster critical AI literacy among SDL learners and teachers.

**Implications for Practice**

Our findings have implications for practice and research on teaching and learning. From an institutional perspective, there is a requirement for frameworks, training, and guidance on how to best incorporate GenAI into SDL. To this end, we note that to date, no studies have addressed the inbuilt biases, cultural orientation, and equity concerns that are part of AI in education (Holmes & Tuomi, 2022; Roe, 2024). Furthermore, while personalisation has been said to form a part of the narrative on AI in education, and is one of the major themes in our review, there is currently no consensus on what this means (Holmes & Tuomi, 2022).

Our findings also suggest that educators' evolving role in GenAI-assisted SDL environments is important. If the role of a teacher in an SDL environment is to shift even further toward a facilitator of GenAI tools to assist the learner in driving their own process and selecting their own goals, using GenAI to help, then this represents a dramatic shift in the current role of the teacher, particularly those who are not frequent users of GenAI tools or are unfamiliar with how they work. In this regard, there is a need for professional development programs and training interventions for practitioners planning to participate in GenAI-enabled SDL. For students, further research is required on the extent to which the benefits of GenAI truly materialise in terms of learning experience and learning outcomes. Consequently, longitudinal and ethnographic research may provide insights into how these technologies benefit students studying an SDL program.

## Conclusion

This scoping review of GenAI and SDL sought to map the existing body of knowledge in this evolving and rapidly developing area. Our synthesis of the current research led to the development of four themes: GenAI as a potential enhancement for SDL, The educator as a GenAI guide, personalisation of learning, and approaching with caution. Our findings suggest that while there are potential opportunities for GenAI to help learners in SDL contexts and provide support, interaction, and potentially personalised learning, there is not yet enough empirical evidence to create a clear roadmap for implementing GenAI tools in this process. At the same time, existing research tends to focus on LLMs and chatbots, with a tendency towards ChatGPT. In the future, further focus will be needed on emerging forms of GenAI, such as NotebookLM (Google, 2024) and other forms of multimodal GenAI. Our identified themes suggest that, at present, GenAI is viewed as a boon to SDL and a tool which may be able to revolutionise the process. At the same time, there is some critical engagement with the risks of these tools, including their unreliability, potential for bias, and the need for critical AI literacy. Our analysis contributes to the field by providing a comprehensive overview of the current research while pinpointing key areas for future investigation.

The confluence of GenAI and SDL practices represents a challenging task. Success will depend on maintaining a balanced, informed, and research-backed approach that encourages the use of GenAI in contexts and cases where the benefits outweigh the potential trade-offs. Ultimately, the success of GenAI in an SDL context will be determined by our ability to distinguish between these scenarios, in which we are able to utilise and understand the benefits of certain technologies in learning, while not ascribing to the belief that technology will by default enhance the learning experience.

## Statements and Declarations

**Funding**

No funding was received for this research work.

**Competing Interests**

The authors declare no competing interests.

**Data Availability Statement**

All data used in this scoping review was obtained from publicly available research articles and publications. These publications are listed in the reference section, and the complete search strategy, inclusion criteria, and data extraction tables are available from the corresponding author upon reasonable request.